# A note on Tesla's revised safety report crash rates

Noah J. Goodall[1]*

[1]CET Research, LLC, the United States of America

*Corresponding author: noah@cetresearch.com

**Abstract:** Between June 2018 and December 2023, Tesla released quarterly safety reports citing average miles between crashes for Tesla vehicles. Prior to March 2021, crash rates were categorized as (*1*) with their SAE Level 2 automated driving system Autopilot engaged, (*2*) without Autopilot but with active safety features such as automatic emergency braking, and (*3*) without Autopilot and without active safety features. In January 2023, Tesla revised past reports to reflect their new categories of with and without Autopilot engaged, in addition to making small adjustments based on recently discovered double counting of reports and excluding previously recorded crashes that did not meet their thresholds of airbag or active safety restraint activation. The revisions are heavily biased towards no-active-safety-features—a surprising result given prior research showing that drivers predominantly keep most active safety features enabled. As Tesla's safety reports represent the only national source of Level 2 advanced driver assistance system crash rates, clarification of their methods is essential for researchers and regulators. This note describes the changes and considers possible explanations for the discrepancies.

## 1  Background

The safety of automated driving systems (ADSs) is a major research focus. Automated driving systems are generally classified based on the amount of assistance provided using SAE's taxonomy of levels of automation (SAE, 2021). Level 2 refers to automated driving systems that combine both latitudinal and longitudinal control but require continuous oversight by a human driver. An example of Level 2 automation is simultaneous use of lane centering technology and adaptive cruise control, such as Tesla's Autopilot feature. Levels 3–5 differ in that they do not require continuous human oversight in at least certain scenarios.

There have been multiple studies evaluating the crash rates of both Level 2 systems (Goodall, 2024; Leslie et al., 2022; HLDI, 2017; Quality Control Systems, 2019) and Level 3 systems (Blanco et al., 2016; Dixit et al., 2016; Goodall, 2021; Schoettle & Sivak, 2015; Teoh & Kidd, 2017). A significant challenge in determining crash rates of ADS is a lack of data, as police crash records do not report whether vehicles were running ADS at the time of the crash. In an attempt to generate additional crash data, the National Highway Traffic Safety Administration (NHTSA) in June 2021 issued a Standing General Order requiring vehicle manufacturers to report all known severe Level 2 crashes and all known crashes at Levels 3–5, but did not require manufacturers to report automated mileages (NHTSA, 2021). In the absence of exposure data such as miles traveled or hours of operation, researchers cannot calculate or compare crash rates from NHTSA data (Goodall, 2023b).

Another challenge in evaluating ADS crash rates is determining appropriate benchmark crash rates. National crash rate data is often incomparable due to differences in driving environment (Goodall, 2024), vehicle age (NHTSA, 2013), and crash severity definitions (Scanlon et al., 2023). Establishing a fair, 'apples-to-apples' benchmark requires controlling for these factors without introducing additional confounding factors. This has proved challenging, with most benchmarks suffering from inconsistent crash definitions and underreporting of general public crashes (Scanlon et al., 2023).

Without independent or government crash rate data, researchers are forced to rely on manufacturers' voluntarily-reported crash rates when evaluating ADS safety. As these figures are often aggregated over time, space, and vehicle model-year, it is essential that researchers carefully consider data quality as well as contextual clues that may help normalize crash rates to the general driving population (Goodall, 2024).

## 1.1 Tesla safety reports

Tesla Inc. is one of the largest manufacturers of electric vehicles in the world, with 1.3 million vehicles delivered in 2022 [Tesla (2023a), p. 3]. Their SAE Level 2 ADS, Autopilot, has been offered as a feature since 2015 and has accumulated over 3 billion miles of on-road driving (Karpathy, 2020).

Tesla has issued quarterly safety reports since the third quarter of 2018. Tesla uses a standard calendar, with Q1 corresponding to 1 January through 31 March. Tesla's safety reports represent a promising data source where crash data are collected for vehicles both with and without Level 2 automation using a consistent, albeit non-standard, severity threshold using a vehicle population with similar ages, body styles, and safety features. While driving environment differs greatly— prior to November 2022, Level 2 automation was used almost exclusively on freeways (Nedelea, 2022)— this can somewhat be controlled for using data from naturalistic driving studies (Goodall, 2024).

Initially, Tesla reported the average million miles between crashes for three operating categories (Tesla, 2022):

- 'Autopilot engaged'
- 'without Autopilot but with our active safety features'
- 'without Autopilot and without our active safety features'.

Beginning in Q2 2021, Tesla began using different categories (Tesla, 2022):

- 'Autopilot technology (Autosteer and active safety features)'
- 'not using Autopilot technology (no Autosteer and active safety features)'.

Initially, Tesla used these new categories for Q2 2021 and later, while keeping the old categories for Q1 2021 and prior. These changes coincided with NHTSA's Standing General Order of June 2021, although it is unclear if the timing was intentional or coincidental on Tesla's part.

According to the Internet Archive's Wayback Machine (Ogden et al., 2023), sometime between 5 January (Tesla, 2023f) and 8 January 2023 (Tesla, 2023g), Tesla revised all safety reports to match the new two-category system, but changed the category descriptions slightly:

- 'using Autopilot technology'
- 'not using Autopilot technology'.

In addition to changing the category names, Tesla also revised their crash rates, even for those with identical definitions such as 'Autopilot engaged' and 'using Autopilot technology'. In the footnotes, Tesla clarified that they had '*discovered reports of certain events where no airbag or other active restraint deployed, single events that were counted more than once, and reports of invalid or duplicated mileage records rates*' (Tesla, 2023f).

While the original crash rates are no longer listed on Tesla's website, rates between Q3 2018 and Q1 2021 were recorded in the literature (Goodall, 2024), and all rates were recorded by the Internet Archive's Wayback Machine (Tesla, 2024), a digital archive of web content (Ogden et al., 2023). The remainder of this note investigates how the crash rates changed as a result of the January 2023 revision.

## 2 Analysis

### 2.1 Autopilot crash rates

Many quarters showed substantially different Autopilot crash rates postrevision, from a 10.8% reduction in crashes per mile in Q2 2020 to a 12.7% increase in Q3 2019 compared to crash rates prior to the revision. The average change was a 3.6% reduction in Autopilot crashes. Prerevision and postrevision Autopilot crash rates are shown in Table 1.

**Table 1** Autopilot crash rates from Tesla safety reports

| Quarter | Crashes per 100 million miles | | Difference |
|---|---|---|---|
| | Prerevision | Postrevision | |
| Q3 2018 | 29.9 | 29.9 | -0.3% |
| Q4 2018 | 34.4 | 35.2 | 2.5% |
| Q1 2019 | 34.8 | 34.7 | -0.3% |
| Q2 2019 | 30.6 | 30.9 | 0.9% |
| Q3 2019 | 23.0 | 26.0 | 12.7% |
| Q4 2019 | 32.6 | 32.2 | -1.3% |
| Q1 2020 | 21.4 | 20.6 | -3.7% |
| Q2 2020 | 22.1 | 19.7 | -10.8% |
| Q3 2020 | 21.8 | 19.6 | -9.8% |
| Q4 2020 | 29.0 | 26.6 | -8.2% |
| Q1 2021 | 23.9 | 21.6 | -9.7% |
| Q2 2021 | 22.7 | 20.2 | -10.7% |
| Q3 2021 | 20.1 | 18.1 | -10.3% |
| Q4 2021 | 23.2 | 23.0 | -0.9% |
| Average | 26.4 | 25.6 | -3.6% |

### 2.2 Non-Autopilot crash rates

The second category, 'not using Autopilot technology' (hereafter referred to as 'non-AP') implies that any crash not in the first category must be in the second category. An analysis of the data, however, raises several questions.

Table 2 and Figure 1 show crash rates for all non-Autopilot categories visible on Tesla's website on 28 December 2022 (prior to the revision) and 24 February 2024 (after the revision).

Note that 'not using Autopilot technology' tracks closely with the two categories 'without Autopilot and without our active safety features' and 'not using Autopilot technology (no Autosteer and active safety features)'. This suggests that the new revised category is heavily weighted toward crashes and miles where active safety systems were not running.

This is surprising, as studies have shown that drivers generally leave most active safety features running. The Highway Loss Data Institute (HLDI) conducted a study in 2016 at vehicle service departments (Reagan et al., 2018). Vehicles arriving for service were checked to see if various driver assistance systems were activated. The authors found that 93% of the 659 front crash prevention systems were turned on. HLDI's study was limited to vehicles where the on/off setting was maintained through multiple ignitions cycles, and excluded vehicles where features were reverted to a default status upon ignition. These results indicate that a high percentage of drivers use active safety features such as front crash prevention.

Many Tesla vehicles are equipped with four active safety features: lane assist, collision avoidance assist, speed assist, and cabin camera. Collision avoidance assist consists of several sub-features, one of which is automatic emergency braking. According to the most recent owner's manuals of the Model S, X, Y, and 3 (software version 2023.32), automatic emergency braking is activated by default at the beginning of every ignition cycle, and operates continuously when the vehicle travels at speeds between approximately 3 mi/hr (5 km/hr) and 124 mi/hr (200 km/hr) [Tesla (2023c), p. 136; Tesla (2023d), p. 154; Tesla (2023e), p. 141; Tesla (2023b), p. 128]. Automatic emergency braking can only be deactivated by shifting the vehicle into park and selecting Controls > Autopilot > Automatic emergency braking from the vehicle's menu. This procedure must be repeated at the beginning of each drive, as automatic emergency braking is active by default.

Tesla vehicles appear to have at least one form of active safety systems, automatic emergency braking, running by default across nearly all speeds and driving conditions. Even though drivers can deactivate this feature at the beginning of every drive by pushing three or four buttons from the menu, studies show that only 7% actually do so (Reagan et al., 2018), even when their cars remember their selection trip-to-trip (Tesla's do not). This suggests that for nearly all non-Autopilot miles, an active safety feature is running.

If active safety features are active on nearly all Tesla vehicle miles, then one would expect the non-AP crash rate to be very close to the 'without Autopilot but with our active safety features' crash rate from the prerevision reports, as these would represent the vast majority of all non-AP mileage. Analysis shows that this is in fact not the case, as the revised non-AP crash rate is much closer to the 'without Autopilot and without our active safety features' crash rate. The next section discusses potential reasons for this discrepancy.

**Table 2** Non-Autopilot crash rates from Tesla safety reports

| Quarter | Non-Autopilot crashes per 100 million miles | | | |
|---|---|---|---|---|
| | 'not using Autopilot technology (no Autosteer and active safety features)' prerevision | 'with our active safety features' prerevision | 'without our active safety features' prerevision | 'not using Autopilot technology' postrevision |
| Q3 2018 | - | 52.1 | 49.5 | 52.1 |
| Q4 2018 | - | 63.3 | 80.0 | 81.3 |
| Q1 2019 | - | 56.8 | 79.4 | 80.6 |
| Q2 2019 | - | 45.7 | 70.9 | 71.4 |
| Q3 2019 | - | 37.0 | 54.9 | 64.1 |
| Q4 2019 | - | 47.6 | 61.0 | 67.6 |
| Q1 2020 | - | 44.1 | 64.1 | 61.3 |
| Q2 2020 | - | 50.3 | 70.4 | 69.0 |
| Q3 2020 | - | 41.3 | 55.9 | 56.2 |
| Q4 2020 | - | 48.8 | 78.7 | 79.4 |
| Q1 2021 | - | 48.8 | 102.2 | 101.6 |
| Q2 2021 | 83.3 | - | - | 73.5 |
| Q3 2021 | 62.5 | - | - | 63.3 |
| Q4 2021 | 62.9 | - | - | 65.8 |
| Q1 2022 | - | - | - | 82.6 |
| Q2 2022 | - | - | - | 64.9 |
| Q3 2022 | - | - | - | 58.5 |
| Q4 2022 | - | - | - | 71.4 |
| Q1 2023 | - | - | - | 90.9 |
| Q2 2023 | - | - | - | 68.5 |
| Q3 2023 | - | - | - | 65.8 |
| Q4 2023 | - | - | - | 100.0 |
| Mean | 69.6 | 48.7 | 69.7 | 72.3 |
| Std. Dev. | 11.9 | 7.2 | 15.1 | 13.1 |
| Coeff. of Variation | 17.1% | 14.7% | 21.6% | 18.1% |

## 3 Discussion

There are at least three potential reasons why Tesla's non-AP crash rate more closely matches the 'without Autopilot and without our active safety features' crash rate despite evidence that active safety features run for the majority of driving. This section evaluates the likelihood of each explanation.

### 3.1 Universal quantification

In predicate logic, universal quantification asserts that a property must hold true for all members of its domain (Shapiro & Kouri Kissel, 2022). When Tesla stated that crash rates were for vehicles running 'without Autopilot but with our active safety features' (Tesla, 2023a), they may have been using a universal quantification with regards to active safety features, and thereby including crashes and miles that occurred only when *all* active safety feature were engaged. If the vast majority of crashes and miles occurred when at least one active safety system was either disabled by the driver or outside of the system's design domain, then the postrevision crash rates would be biased towards 'without active safety feature' crash

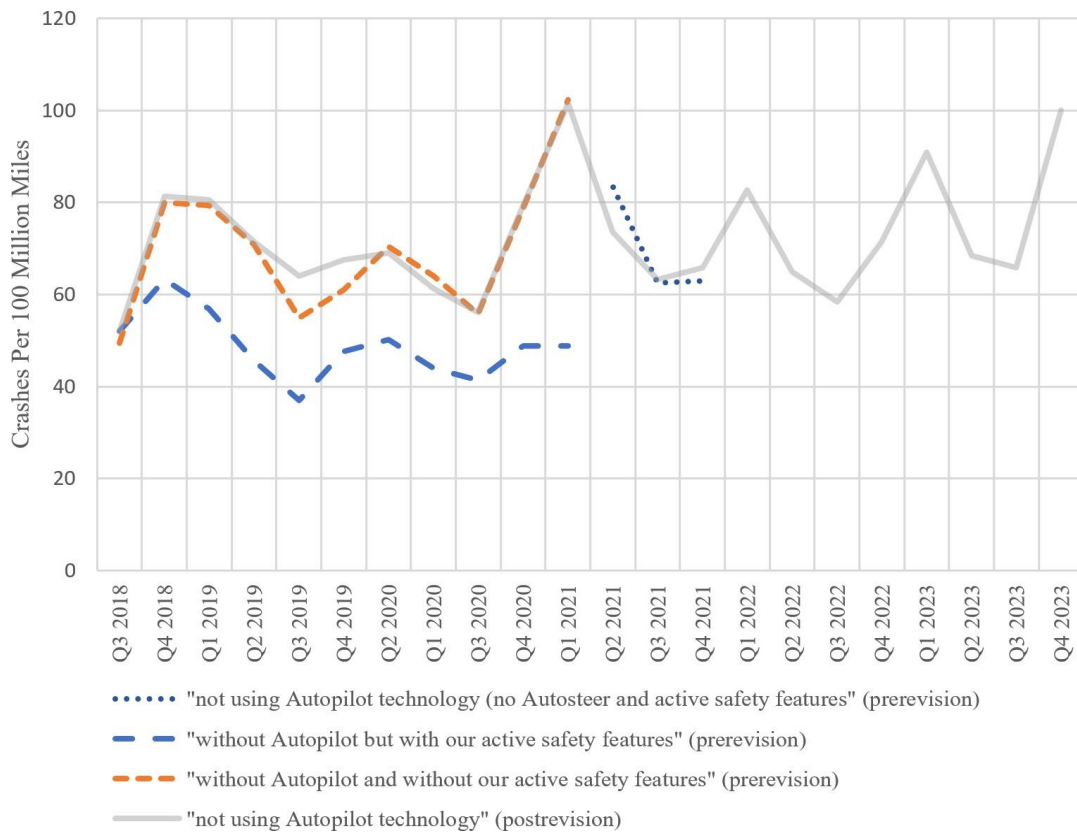

**Figure 1** Non-Autopilot crash rates from Tesla safety reports

rates, as was observed in the revised data.

Tesla's safety manuals, however, show that is impossible for all active safety features to run simultaneously. For example, Obstacle-aware acceleration, a sub-feature of Collision avoidance assist, is designed only to work at speeds between 0 and 10 mi/hr (16 km/hr) [Tesla (2023c), p. 137], while Lane departure avoidance, a sub-feature of Lane assist, is only operational when driving between 40 and 90 mi/hr (64 and 145 km/hr) 'on roads with clearly visible lane markings' [Tesla (2023c), p. 132]. Neither system can operate at speeds between 10 and 40 mi/hr, and so universal quantification does not explain the observed bias in the revised rates. Table 3 shows the operational speeds of various Tesla active safety features in software version 2023.32.

### 3.2 Other classification schemes

Tesla may be using a different method to classify non-AP crashes as either with or without applicable active safety features. Although there is no way to determine this from the available data, some possible schemes are:

- **Exclusion of certain applicable active safety features.** Tesla may be using a universal quantification but excluding specific features, such as Obstacle Aware Acceleration. This would permit all other active safety features to be running in certain instances, thereby classifying them as a small sample of 'with active safety features' crashes and miles.

- **Different minimum feature thresholds.** Tesla may instead consider a crash to be 'with active safety features' if two, three, four, etc. features are running simultaneously.

- **Differences in model-year vehicles.** Active safety features were reviewed for a recent software version in all four Tesla models. There may be earlier software versions on Tesla vehicles with fewer active safety features or with different operational speed or other characteristics that may result in a large percentage of vehicles operating without active safety features per Tesla's definition. Given the numerous software versions, possible variants in crash classification schemes, and uncertainty in crash/mileage sample size, a detailed review of vehicle specifications is unlikely to resolve these questions and was considered outside the scope of this note.

**Table 3** Tesla safety features, software version 2023.32

| Active safety feature | Speed range | Notes |
| --- | --- | --- |
| **Lane assist (without Autosteer)** | 7–85 mi/hr (12–140 km/hr) | approximate |
| Steering interventions | 30–85 mi/hr (48–140 km/hr) | 'on major roads with clearly visible lane markings' |
| Lane departure avoidance | 40–90 mi/hr (64–145 km/hr) | 'on roads with clearly visible lane markings' |
| Emergency lane departure avoidance | 40–90 mi/hr (64–145 km/hr) | 'on a road with clearly visible lane markings, curbs, etc.' |
| Blind spot assist | 0–140 mi/hr (0–240 km/hr) | inferred from manuals |
| Automatic blind spot camera | 0–140 mi/hr (0–240 km/hr) | inferred from manuals |
| Blind spot collision warning chime | 0–140 mi/hr (0–240 km/hr) | inferred from manuals |
| **Collision avoidance assist** | | |
| Forward collision warning | 3–124 mi/hr (5–200 km/hr) | approximate |
| Automatic emergency braking | 3–124 mi/hr (5–200 km/hr) | approximate |
| Multi-collision braking | 0–140 mi/hr (0–240 km/hr) | inferred from manuals |
| Obstacle-aware acceleration | 0–10 mi/hr (0–16 km/hr) | |
| **Speed assist** | 0–140 mi/hr (0–240 km/hr) | inferred from manuals |
| **Cabin camera** | 0–140 mi/hr (0–240 km/hr) | inferred from manuals |

There may be other classifications schemes with consistent logic and utilization at play. Additional context would benefit researchers.

### 3.3 Unreported data

It is possible that some data was not included in the revised crash rates, either as part of the exclusions mentioned in the January 2023 update (Tesla, 2023f) or by some other means. This theory has some support under the terminology used for the Q2–Q4 2021 reports, where 'not using Autopilot technology (no Autosteer and active safety features)' could be read as 'no Autosteer and no active safety features'. Under this interpretation, the crashes involving active safety features without Autosteer would not have been counted.

Additional evidence can be found by analyzing the relative weights of combined averages of crash rates. Before the revisions, with and without active safety features crash rates are reported separately, while after the revisions they are reported together. Although precise mileage and crash counts are unavailable, the relative weights of the two prerevision crash rates in the combined postrevision crash rate can be calculated using the formula for average of averages:

$$\mu = \frac{n_1\mu_1 + n_2\mu_2}{n_1 + n_2}. \quad (1)$$

Substituting $n_2 = 1 - n_1$ yields:

$$n_1 = \frac{\mu - \mu_2}{\mu_1 - \mu_2}. \quad (2)$$

The sample sizes $n_1$ and $n_2$ represent the relative sample sizes of prerevision crash rates in the postrevision crash rate $\mu$.

In 8 of the 11 quarters, the revised crash rate is higher than both of the prerevision crash rates, resulting in negative weights. This is probably due to reclassification of crashes as acknowledged by Tesla in the revision. In the remaining three quarters, the crashes with active safety features represented 7% (Q2 2020), 14% (Q1 2020), and 1% (Q1 2021) of the revised crash rate. Focusing on the most balanced quarter, Q1 2020, 'without active safety features' represented 86% of the revised crash rate while 'with active safety features' represented 14%. As crash rate is defined as crashes per vehicle-miles traveled (and assuming insignificant changes to the number of crashes and miles as part of the revision), then non-Autopilot mileage without active safety features was approximately 6.3 times greater than with active safety features for that quarter. It seems highly unlikely that vehicles traveled 6.3 times further without active safety features than with given that active safety features are activated by default at each ignition cycle.

According to the central limit theorem, as the sample size of a population of means increases, the sample mean variance will decrease (Kwak & Kim, 2017). When applied to Tesla crash rates, and assuming vehicles without active safety features traveled at least 6.3 times further than vehicles with active safety features, one would expect the 'without active safety features' crash rates to be more consistent over time than those of vehicles with active safety features due to their larger sample sizes.

Because the quarterly crash reports represent different Tesla software versions in a continuously evolving vehicle fleet, these cannot be considered means from the same population and so a direct analysis of sample mean variance cannot be performed. Surrogate techniques, however, suggest that the supposed larger sample size yields greater variability. First, a visual inspection of Figure 1 does not show an observable difference in variation. Second, the sample standard deviation of 'with active safety features' is 7.2 (14.7% coefficient of variation), lower than 'without active safety features' at 15.1 (21.6% coefficient of variation). This is a counterintuitive result, given that 'without active safety features' was expected to have a larger sample size based on the analysis of combined averages, and therefore a smaller coefficient of variation.

The limited evidence suggests that 'with active safety features' crashes may have actually been the larger sample before the revisions as indicated by the smaller coefficient of variation and the general prevalence of active safety features on Tesla vehicles. Tesla's crash rate revisions, however, suggest that 'with active safety features' crashes were the smaller sample. Simply reclassifying crashes from 'with active safety' to 'without' would not affect the combined average. Instead, the most efficient way to achieve these results would be to remove or otherwise undercount a portion of the 'with active safety features' crashes.

This cursory analysis raises several questions about sample size, crash classifications, and data transparency that require additional context before any conclusions can be drawn regarding the safety and crash rates of Tesla vehicles.

## 4 Conclusions

This note examined Tesla's revised safety reports covering the period from July 2018 to December 2023, which document the average miles between crashes for Tesla vehicles. Crash rates were reported as three categories prior to March 2021: Autopilot, active safety features only, and neither. In January 2023, Tesla revised all historical crash rates to a two-category system: 'using Autopilot technology' and 'not using Autopilot technology'. Tesla further updated crash rates to reflect recently discovered instances of double counting and false positives.

The analysis presented as part of this note revealed counterintuitive crash rates, particularly in the 'not using Autopilot technology' category where the data suggested that crashes without active safety features were heavily weighted in the postrevision reports. This finding is surprising, as prior research suggests that drivers usually keep active safety features enabled.

Several potential explanations for this discrepancy were discussed, including the possibility that Tesla used a universal quantification approach, which includes all crashes and miles that occurred when all active safety features were engaged. However, analysis showed that it is impossible for all active safety features to run simultaneously due to differing operational speeds and conditions. Thus, this explanation is unlikely to account for the observed results.

Another possibility is that Tesla employed an alternate classification scheme that excluded specific features or applied different minimum feature thresholds when categorizing crashes. While this is plausible, more context and information are needed to verify this hypothesis.

Finally, the possibility of unreported data or exclusions in the revised crash rates was considered. Some evidence suggests that certain data might not have been included in the revised rates, potentially affecting the overall results. In order to produce the revised crash rates when combining with and without active safety features crashes, vehicles would need to have traveled at least 6.3 times further with no active safety features active. This is unlikely, as default active safety features are rarely deactivated by individual drivers. Furthermore, the supposed smaller sample sizes of 'with active safety features' do not exhibit smaller variances as would be expected.

In conclusion, Tesla's revisions to its safety reports raise questions and warrant further investigation. The discrepancies observed in the crash rate data, especially in the 'not using Autopilot technology' category, highlight the need for greater transparency and clarity in reporting methodologies. Given the limited data on automated driving system crash rates, a clearer understanding of the data sources and classifications used by manufacturers like Tesla is essential for researchers, regulators, and consumers.

## CRediT contribution statement

**Noah J. Goodall:** Conceptualization, Formal analysis, Investigation, Methodology, Visualization, Writing—original draft, Writing—review & editing.

## Declaration of competing interests

The author has served as a paid expert witness for the plaintiffs in litigation filed against Tesla.

## Funding

This work did not receive external funding.

## Acknowledgements

Justice Appiah, Michael Fontaine, Geoff Keeling, and Megan Beth Koester reviewed draft sections of this paper.

An earlier version of this article was published as a preprint on arXiv on 10 November 2023 (Goodall, 2023a).

## About the authors

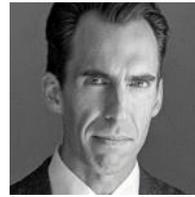

**Noah J. Goodall** received the Ph.D. degree in civil engineering from the University of Virginia, Charlottesville, VA, USA, in 2013. He is a researcher with CET Research, LLC. His research interests are vehicle automation, intelligent transportation systems, crowdsourced data, and safety.

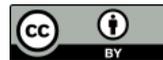